\documentclass[twocolumn, a4paper, 11pt,book]{archive}
\usepackage{lineno,hyperref}
\modulolinenumbers[5]
\modulolinenumbers[1]
\usepackage{color}
\usepackage{multirow,bigdelim}
\usepackage{version}
\usepackage{graphicx}

\usepackage[]{natbib}

\usepackage{txfonts}

\usepackage{booktabs}
\usepackage{multirow}

\usepackage{rotating} 
\usepackage{color, colortbl}

\usepackage{siunitx}

\definecolor{Gray}{gray}{0.9}
\definecolor{LightRed}{rgb}{1., 0.88, 0.88}
\definecolor{LightCyan}{rgb}{0.88,1,1}
\definecolor{LightGreen}{rgb}{0.8,0.98,0.8}
\definecolor{LightViolet}{rgb}{0.95, 0.85, 1.0}
\definecolor{azur}{rgb}{0.94, 1.0, 1.0}

\begin{document} 

\twocolumn[{%
 \centering
%
{\center \bf \Huge Carbon dioxide clathrate hydrate formation at low temperature.}\\
{\center \bf \Large Diffusion-limited kinetics growth as monitored by FTIR}\\
\vspace*{0.25cm}

{\Large E. Dartois \inst{1},
        F. Langlet\inst{2},
 }\\
\vspace*{0.25cm}

$^1$      Institut des Sciences Mol\'eculaires d'Orsay, CNRS, Universit\'e Paris-Saclay, 
B\^at 520, Rue Andr\'e Rivi\`ere, 91405 Orsay, France\\
              \email{emmanuel.dartois@universite-paris-saclay.fr}\\
$^2$      Institut d'Astrophysique Spatiale (IAS), UMR8617, CNRS, Universit\'e Paris-Saclay, B\^at. 121, 91405 Orsay, France\\
 \vspace*{0.5cm}
{keywords: Infrared: planetary systems; Planets and satellites: composition; Comets: general; Methods: laboratory: solid state; Solid state: volatile; Molecular processes}\\
 \vspace*{0.5cm}
{\it \large To appear in Astronomy \& Astrophysics}\\
 \vspace*{0.5cm}

%
%
 }]
%
  \section*{Abstract}
  {The formation and presence of clathrate hydrates could influence the composition and stability of planetary ices and comets; they are at the heart of the development of numerous complex planetary models, all of which include the necessary condition imposed by their stability curves, some of which include the cage occupancy or host-guest content and the hydration number, but fewer take into account the kinetics aspects.
   We measure the temperature-dependent-diffusion-controlled formation of the carbon dioxide clathrate hydrate in the 155-210~K range in order to establish the clathrate formation kinetics at low temperature.
   We exposed thin water ice films of a few microns in thickness deposited in a dedicated infrared transmitting closed cell to gaseous carbon dioxide maintained at a pressure of a few times the pressure at which carbon dioxide clathrate hydrate is thermodynamically stable.
The time dependence of the clathrate formation was monitored with the recording of specific infrared vibrational modes of CO$_2$ with a 
Fourier Transform InfraRed (FTIR) spectrometer.
   These experiments clearly show a two-step clathrate formation, particularly at low temperature, within a relatively simple geometric configuration. We satisfactorily applied a model combining surface clathration followed by a bulk diffusion-relaxation growth process to the experiments and derived the temperature-dependent-diffusion coefficient for the bulk spreading of clathrate.
The derived apparent activation energy corresponding to this temperature-dependent-diffusion coefficient in the considered temperature range is $\rm E_a =$24.7 $\pm$ 9.7 kJ/mol. 
The kinetics parameters favour a possible carbon dioxide clathrate hydrate nucleation mainly in planets or satellites.}
%

%

\section{Introduction}
\label{intro}

Clathrate hydrates are inclusion compounds that  trap molecules in a crystalline water ice network. For the two main structures, these clathrate hydrates form water-based cages of different sizes and of two types, a larger one (forming a hexagonal truncated trapezohedron or hexadecahedron) and a smaller one (forming a pentagonal dodecahedron), leading to the simplest clathrate structures I and II, respectively. 
%
\begin{figure*}[htbp]
\begin{center}
\includegraphics[width=1.8\columnwidth,angle=0]{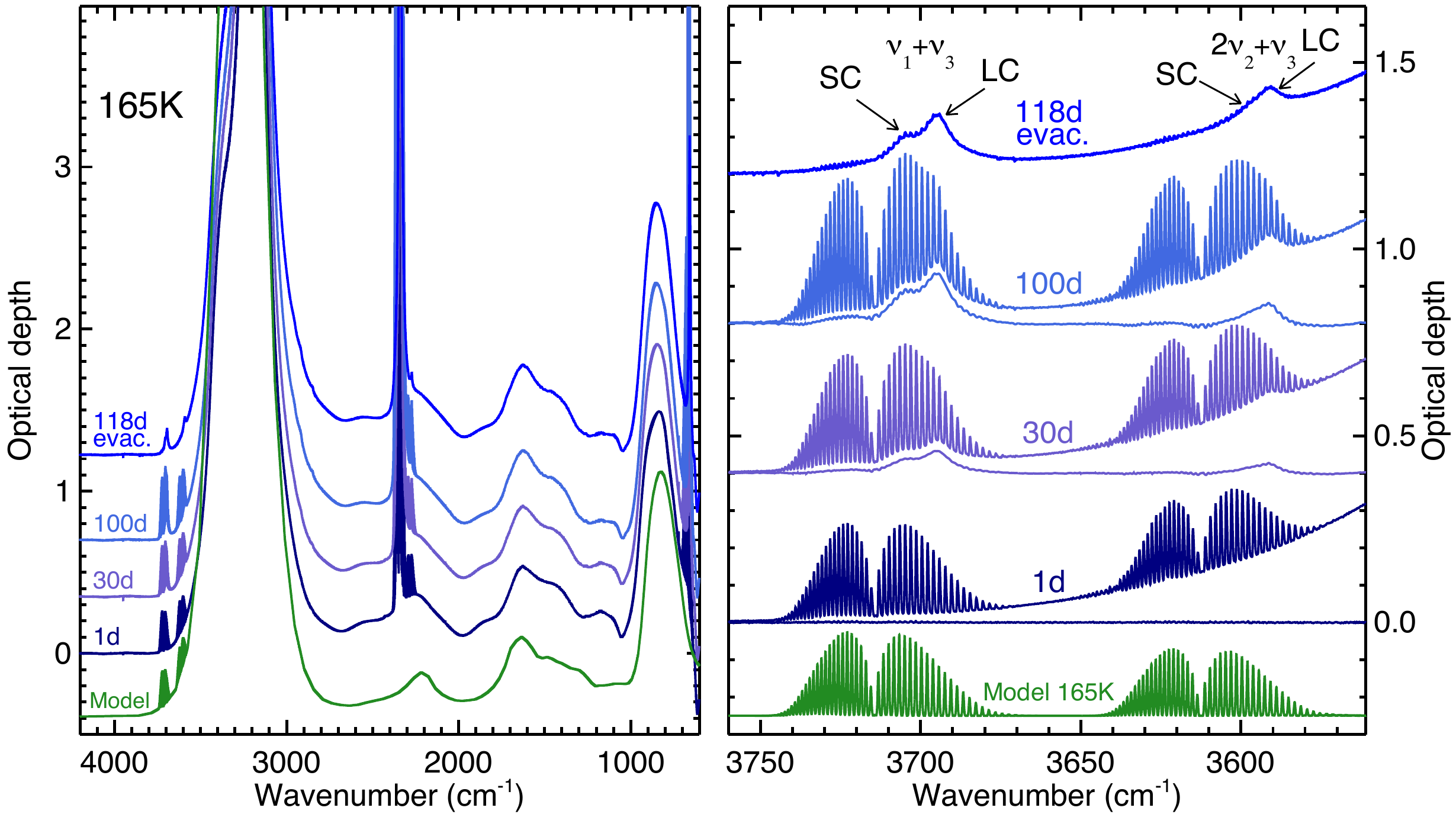}
\caption{Infrared spectra of the clathrate kinetics experiment performed at 165K. Spectra extracted after 1, 30, and 100 days of exposure of the ice film to CO$_2$ gas at 60 mbar are shown in the 4500-600 cm$^{-1}$ range in the left panel. The upper trace is the spectrum recorded just after evacuation of the gas at the end of the experiment after 118 days. A close-up on the gas and clathrate hydrate signatures in the $\rm \nu_1+\nu_3$ and $\rm 2\nu_2+\nu_3$ combination modes is shown in the right panel. Under each spectrum is shown the baseline-corrected spectrum of the contribution by the CO$_2$ in the clathrate cages after subtraction of the gas phase contribution to the spectrum.
The CO$_2$ signatures in the small (SC) and large (LC) cages of the clathrate hydrate are labelled in the upper trace. The lower trace is a model of the CO$_2$ gas phase at 165K. See text for details.}
\label{Figure_explicative}
\end{center}
\end{figure*}
%
%
%
On Earth, the main interest in clathrate hydrates is focused on methane clathrate hydrate, either as a potential hydrocarbon resource \citep[e.g.][]{Jin2020, Boswell2020} or for for the problems caused by its ability to block pipelines in the petrol industry \citep[e.g.][]{Fink2015} or its participation in climate change if natural fields are destabilised \citep{Riboulot2018}. Carbon dioxide clathrate hydrate is of interest for fundamental research on Earth, or for its potential use in greenhouse gas sequestration \citep{Duc2007, House2006}.

In a planetary or astrophysical context, clathrate hydrates are considered because they can retain volatile molecules at pressures higher than the pure compound that would otherwise sublimate under most astrophysical conditions, and/or influence the geophysics of large bodies \citep[e.g.][]{Fortes2010,Choukroun2010,Bollengier2013}. 
The existence of clathrate hydrates in Solar System bodies has been explored in many models, principally on the basis of the stability curves for these inclusion compounds.
During the evolution of the solar nebula, the existence of clathrate hydrate, or rather its stability, depends on pressure and temperature. These thermodynamic properties represent a necessary but not sufficient condition. In many bodies, in particular small bodies such as comets, or low-temperature objects orbiting at large distances, such as trans-Neptunian objects, including Kuiper Belt objects, the kinetics for their formation becomes a primary driving factor to take into account in the modelling.\\
Carbon dioxide is a significant constituent of large to smaller icy bodies of the Solar System.
The kinetics of carbon dioxide clathrate hydrate formation \citep{Schmitt1986, Schmitt2003, Circone2003, Genov2003} is a key ingredient in any discussion about clathrate formation. Kinetics experiments have been performed on clathrates using, for example, neutron diffraction  experiments monitoring the ice structural change towards clathrate formation \citep{Falenty2013,
Henning2000}, or more classical,
Pressure, Volume, Temperature (PVT) experiments \citep{Falenty2013} in the 272-185K range. These experiments are mostly based on the polydisperse distribution of ice spheres, with peak size distributions generally in the 20-200 microns range.
The interpretation of these data is performed using a modified so-called Johnson-Mehl-Avrami-Kolmogorov (JMAK) phenomenological model and knowledge of the size distribution of ice particles. The duration of the neutron diffraction experiments is  often limited to typically tens of hours because of the availability of diffraction beam lines shifts, which can limit the temperature range explored. Indeed, as the diffusion coefficient generally follows an Arrhenius equation, it evolves exponentially with temperature and going to lower temperatures becomes a difficult issue, requiring long-duration experiments.
In this article, we develop an alternative experimental approach based on the direct measurement, in transmittance, of the infrared signature of the caged molecules within a controlled thin ice film structure. This approach allows us to extend the measurements to lower temperatures (down to 155K) at which the kinetics are measured. The article is organised as follows: in the following section, we describe the kinetics experiments performed and the two-step diffusion and nucleation model adjusted to the data. We then discuss the results, derive the kinetics and apparent activation energy of the low-temperature carbon dioxide clathration, and discuss their astrophysical implications. 
\section{Experiments}
%
\begin{table}[htp]
\caption{Summary of experiments}
\begin{center}
\begin{tabular}{l l l l l}
\hline
T       &P$\rm_{exp}$   &P$\rm_{eq}$$\rm^a$     &Duration       &Thickness      \\
(K)     &{\small(mbar)}         &{\small(mbar)}                                 &h/(days)       &($\mu$m)       \\
\hline
155             &20             &9.32   &430(18)                &2.6$\pm$0.4            \\
165             &60             &23.5   &2820(118)      &3.6$\pm$0.5            \\
170             &120            &37.3   &1260(53)               &0.7$\pm$0.1            \\
177.5   &200            &71.7   &1778(74)               &5.0$\pm$0.7            \\
185             &400            &128.6  &1595(66)               &5.0$\pm$0.7            \\
190             &700            &186.5  &1196(50)               &6.5$\pm$1.0            \\
200             &1200   &365.0  &379(16)                &5.1$\pm$0.7            \\
205             &1600   &498.0  &1080(45)               &20$\pm$3               \\
207.5   &1700   &588.4  &254(11)                &4.9$\pm$0.7            \\
210             &2000   &678.7  &162(7)         &3.0$\pm$0.5            \\
\hline
\end{tabular}
\end{center}
$\rm^a$ From \cite{Miller1970} and \cite{Falabella1975}
\label{summary}
\end{table}
%
%
CO$_2$ was purchased from Air liquide. The initial purity is N25 (99.999\%). 
The experimental setup consists in a gold-coated copper cell, thermally coupled to a liquid N$_2$-transfer cold finger placed in a
high-vacuum, evacuated cryostat ($\rm P < 10^{-7}$mbar). Infrared transmitting zinc selenide windows of 4 mm in thickness facing each other with some space in between, are sealed with indium gaskets, and allow the spectrometer beam to record clathrate hydrate spectra in transmittance. A soldered stainless steel injection tube brazed to the lower part of the cell allows the entry or evacuation of gas. A description of such a cell can be found in \cite{Dartois2010}   for example.
The cell was modified to limit the distance between the two ZnSe windows to a space of the order of 200~$\mu$m, which is sufficiently small to allow us to monitor the clathration of CO$_2$ while the ice film is exposed to the gas.
To prepare the ice film, water vapour is injected into the evacuated cell pre-cooled to 240K, and condenses on the ZnSe windows. The cell is then lowered to the temperature of study.
The guest gas is then injected, with a pressure excess a few times the clathrate stability pressure, and below the CO$_2$ ice condensation pressure. The pressures used are given in Table~\ref{summary}, as well as the expected equilibrium pressure for the clathrate stability.
The thickness of the produced ice film  varied from a fraction of a micron to several microns, and 13~mm in diameter.
The thickness is monitored by Fourier Transform InfraRed (FTIR) spectroscopy, over the central 7~mm of the 
IR windows of the cell.
Spectra were recorded regularly as a function of time, with a spectral resolution set between 0.16 and 0.32~cm$^{-1}$, with a liquid nitrogen (LN$_2$) Mercury Cadmium Telluride (MCT) detector. 
The time dependency sometimes presents some missing data, because the MCT detector, with an autonomy of about 10~h, could not always be kept cold over several months, and therefore no spectra were recorded during some time intervals.

Spectra from the experiment conducted at 165K are shown in Fig.~\ref{Figure_explicative}. These were  recorded at different times after the start of the experiment (1, 30, and 100 days). We also show the spectrum of the clathrate formed recorded immediately after the evacuation of the CO$_2$ gas at the end of the experiment, after 2608 hours of exposure (more than 108 days). The clathrate spectrum is similar to the one expected from the 150K measurement reported in \cite{Dartois2009}, where the small and large  clathrate cage environments were identified, and the structure I clathrate hydrate was confirmed.
A close-up on the $\rm \nu_1+\nu_3$ and $\rm 2\nu_2+\nu_3$ (alternatively named $\rm \nu_1+\nu_3$ Fermi resonances 1 and 2) is shown. These combination modes were selected to follow the evolution of the process of clathration, because the gas phase absorptions are not fully saturated during pressurisation, and the caged CO$_2$ infrared signature can be observed within the comb of the rovibrational transitions, as shown in Fig.~\ref{Figure_explicative}. In addition to the cell temperature diode measurement, the gas phase CO$_2$ temperature can also be monitored via the observed comb. We modelled the gas phase spectrum using the HITRAN database \citep{Kochanov2016} for a 200~$\mu$m path length, a CO$_2$ pressure of 60 mbar, and a gas temperature of 165K, and convolved using the transfer function of the spectrometer. The result is shown in the lower trace in Fig.~\ref{Figure_explicative}, which also confirms the gas thermalisation.
%
%
\begin{figure}[htbp]
\begin{center}
\includegraphics[width=0.9\columnwidth,angle=0]{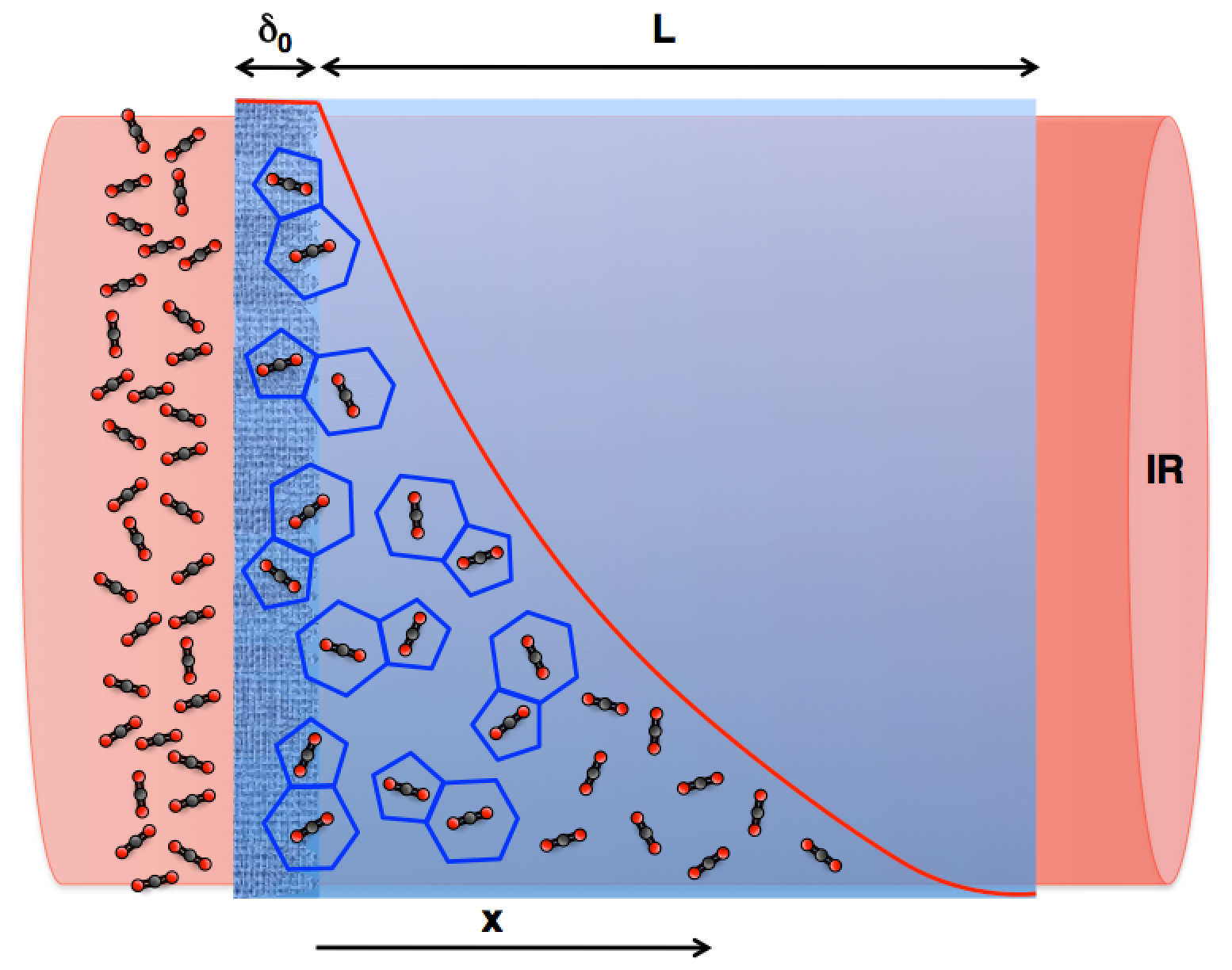}\caption{Schematic representation of the integrated column density measurement in the experiment and dimensions used in equations \ref{JMAK}-\ref{equation_somme}. The gas (on the left) is in contact with the ice film (in blue). The clathration proceeds through the boundary ($\rm \delta_0$), and the bulk clathration and diffusion evolves with time over the length L. The red line  illustrates a snapshot of the clathrate concentration progression in the ice film. The infrared beam (red cylinder) probes the column density of clathrates formed in the entire film at a given time.}\label{schema}
\end{center}
\end{figure}
%
%
\begin{figure}[htbp]
\begin{center}
\includegraphics[width=\columnwidth,angle=0]{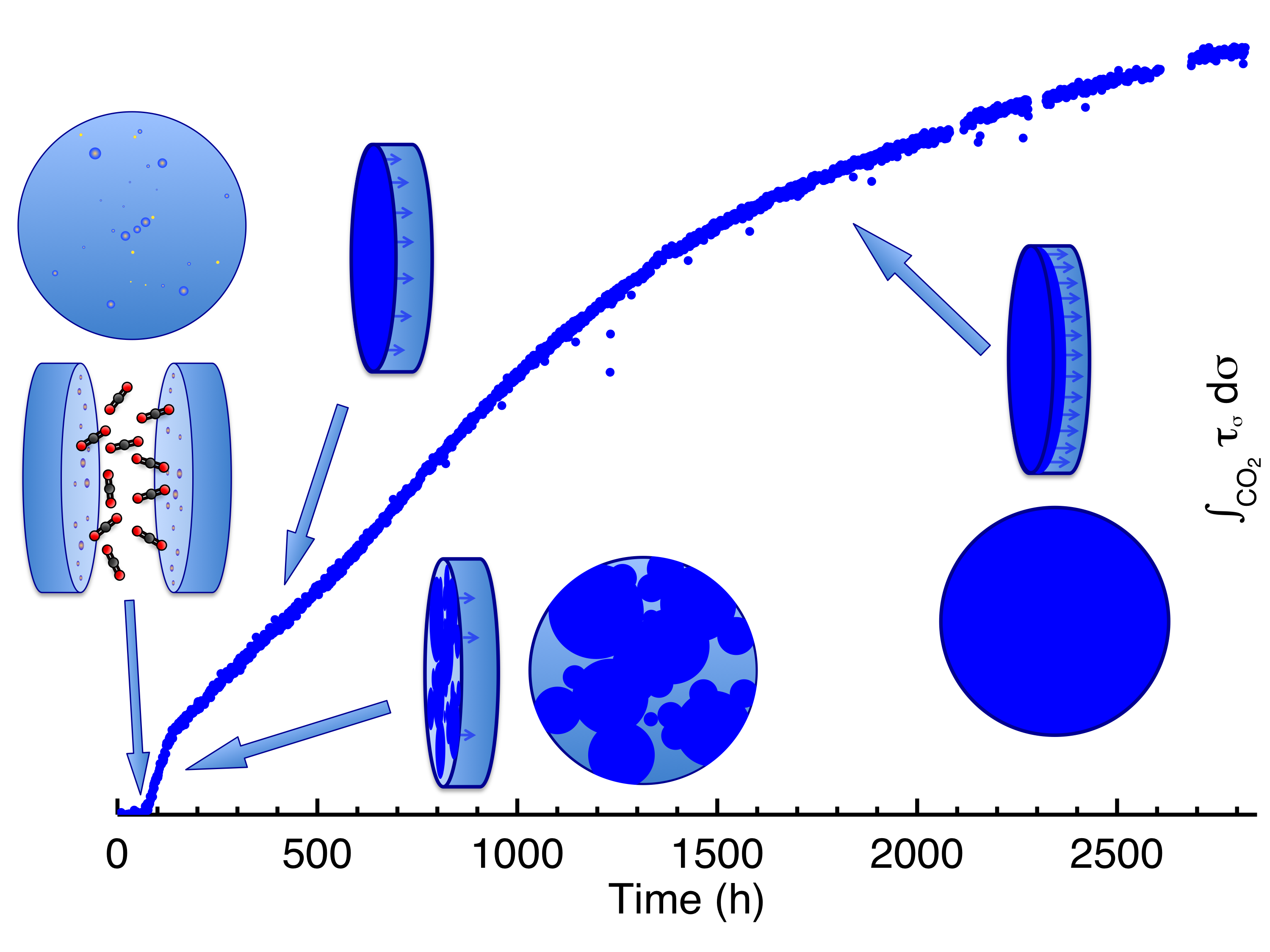}
\caption{Schematic representation of the incubation and nucleation, and diffusion-limited regimes. It is illustrated on the 165K CO$_2$ clathrate time-dependent, integrated optical depth absorption curve evolution, as measured in the experiment.
As time proceeds, we see the evolution of the degree of clathration (blue) of the
ice film (light blue) when exposed to the CO$_2$ gas in the pressurised cell. From left to right, the first inflexion shows the incubation time with an arrow, then JMAK surface clathration spreading and proceeding quickly from clathrate seeds, followed by diffusion-limited regimes of clathrate growth through the interface.
}
\label{schema_evolution}
\end{center}
\end{figure}
%
\section{Model}
The low-temperature experiments show in particular that the clathration of the film is at least a two-step process (already mentioned in other experiments, e.g. \citet{Falenty2013}), which is particularly well observed in the 165 and 175~K experiments where the two steps are well decoupled in the clathration time-sequence. The surface in direct contact with the gas initiates the first phase. The initial phase is kinetically driven and the clathrate surface coverage can be described as points randomly distributed on the ice surface that then grow up to a limited depth $\rm \delta_0$, after which the growth proceeds through a diffusion- or relaxation-limited evolution.
The quasi-two-dimensional evolution of the ice interface  can therefore be described by a Johnson-Mehl-Avrami-Kolmogorov (JMAK) model: a model describing the coverage time dependence and kinetics \citep{Fanfoni1998}, represented by a simple equation:
\begin{equation}
\rm \frac{N_{\scriptstyle JMAK}(t)}{N_{JMAK}(\infty)} = 1 - exp (-Kt^n)\label{JMAK}
,\end{equation}
which characterises the fraction of clathration after a hold time at a given temperature. Here, K and n are constants; n typically varies between 2 and 4 for 2D to 3D transformations; and
$\rm N_{JMAK}(\infty)$ is the column density of clathrate cells, monitored through the CO$_2$ combination modes, corresponding to the characteristic (surface) thickness $\delta_0$.
The second phase occurring after the surface clathration (or concomitantly, at high temperatures) is the diffusion-limited clathrate growth in the bulk of the ice film of thickness L (Fig~\ref{schema}).
The low-temperature evolution of the infrared spectra guides the analysis to be performed, as it is the moment when the different phases are best observed separately.
At late times in the 165K evolution of the CO$_2$ clathrate infrared signature, integrated over the film thickness, the clathrate growth behaviour is purely diffusion-limited.
At intermediate times, it transits from the surface clathration behaviour to a diffusive behaviour (see Fig~\ref{schema_evolution}).
The evolution of the clathrate column density resembles many other so-called two-step diffusion problems, in which the concentration at the surface boundary does not rise immediately to its value at infinite time, either because of the irregular nature of the surface skin already clathrated or because of a kinetic barrier to the solely diffusion-driven clathrate transformation. This kinetic barrier can be influenced by the local CO$_2$ concentration (the transformation can be hindered or much slower at low CO$_2$ diffusive concentrations), or the relaxation needed to accommodate the volume change during clathrate transformation (an empty structure I hydrate would have a density of $\rm 0.796 g/cm^3$ \citep{Uchida1997}, whereas ice Ic and Ih have densities in the range of $\rm 0.916-0.93 g/cm^3$  \citep{Petrenko1999}).
This diffusion behaviour can be modelled using a progressive clathration through the boundary, and the concentration at the boundary becomes time dependent. These variable concentration model conditions can be described by
\begin{equation}
\rm Clh(t) = Clh(t_0) + [Clh(\infty) - Clh(t_0)] [1-exp(-\beta t)]
\label{Long_Richman}
.\end{equation}
At the surface interface, the clathrate hydrate formation proceeds as the concentration rises to a given initial value immediately after the surface clathration ($\rm Clh(t_0)$) and evolves to the final value at equilibrium (infinite time, $\rm Clh(\infty)$), through a first-order relaxation process. {The clathrate concentration thus increases with the difference with respect to the equilibrium value} $\rm Clh(\infty)$, and reaches this equilibrium value  asymptotically at infinite time.
Such a model has been mathematically derived for a membrane of thickness $\rm 2 {\scriptstyle \times} L$, with the two boundaries exposed to the concentration evolution prescribed by equation~\ref{Long_Richman} \citep[e.g.][]{Long1960, Crank1975, Mamaglia2012}. 
Mathematically, the situation of one surface exposed to this concentration evolution and limited by an impermeable boundary at L (the ZnSe substrate surface on which the ice film is deposited in our experiments; c.f. Fig~\ref{schema}) is equivalent to two surfaces of a membrane exposed with no boundary.
The time-dependent solution to this diffusion-limited growth can therefore be described by
\begin{multline}
\rm Clh (\lambda,\theta)=\!\phi \left[ 
1 \! - \! \frac{4}{\pi} \displaystyle\sum_{n=0}^{\infty} 
\frac{(-1)^n}{2n+1} cos \frac{(2n+1)\pi \lambda}{2}
exp\left(\frac{-(2n+1)^2 \pi^2 \theta}{4} \right)
\right]\\
\rm + (1-\phi) \left[ 
1 - \frac{cos(\sqrt{\psi} \lambda) exp(-\psi\theta)}{cos(\sqrt{\psi})}
- \frac{4}{\pi} \displaystyle\sum_{n=0}^{\infty} 
\frac{(-1)^n}{(2n+1) \left(1-\frac{(2n+1)^2\pi^2}{4\psi} \right)} \right.\\ 
\rm \left. cos \frac{(2n+1)\pi \lambda}{2}
exp\left(\frac{-(2n+1)^2 \pi^2 \theta}{4} \right)
\right],
\label{propagation}
\end{multline}
where $\rm \lambda=x/L$, $\rm \phi = Clh(t_0)/Clh(\infty)$, $\rm \theta=D t / L^2$, $\rm \psi = \beta L^2 / D$,
x is the distance from the boundary of the surface clathrate,  L is the film thickness starting from the depth $\delta_0$ of the surface clathrate formation (the total film thickness is thus L+$\delta_0$, see Fig~\ref{schema}), D ($\rm m^2.s^{-1}$) is the diffusion coefficient, and
$\rm \phi$ is the ratio of immediate to long time equilibrium concentrations at the surface clathrate boundary. We note that
$\rm \phi =1$ is the sign of a CO$_2$ immediate diffusion-driven clathration process, in which case the second term of equation~\ref{propagation} vanishes.
The column density evolution of the clathration integrated over the film, i.e. what is measured by infrared spectroscopy, then follows
\begin{multline}
\rm \frac{N_{\scriptstyle D}(t)}{N_{D}(\infty)}= \phi \left[ 1 - \frac{8}{\pi} \displaystyle\sum_{n=0}^{\infty} \frac{exp\left(\frac{-(2n+1)^2 \pi^2 \theta}{4} \right)}{(2n+1)^2} \right] \\
\rm  + (1-\phi) \left[ 1 - \frac{tan\sqrt{\psi}exp(-\psi\theta)}{\sqrt{\psi}}  \frac{8}{\pi} \displaystyle\sum_{n=0}^{\infty} \frac{exp\left(\frac{-(2n+1)^2 \pi^2 \theta}{4} \right)}{(2n+1)^2\left( 1 - \frac{(2n+1)^2 \pi^2}{4\psi}\right)} \right].
\end{multline}
The overall measured column density of the carbon dioxide clathrate formation is the sum of the surface and bulk diffusion-limited growths,
\begin{equation}
\rm N_{total}(t) = N_{\scriptstyle JMAK}(t) + N_{\scriptstyle D}(t).
\label{equation_somme}
\end{equation}
%
\begin{figure}[htbp]
\begin{center}
\includegraphics[width=0.49\columnwidth,angle=0]{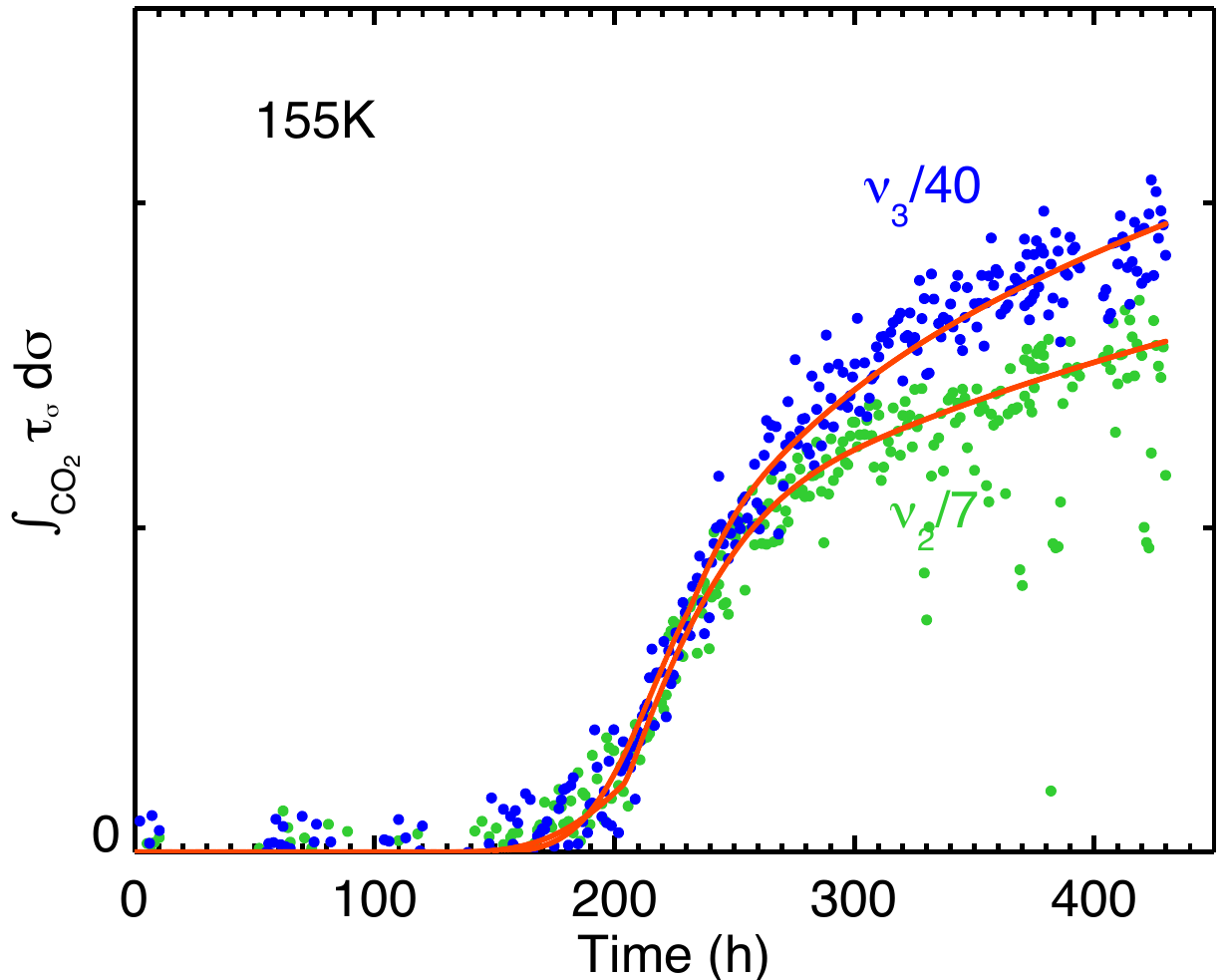}
\includegraphics[width=0.49\columnwidth,angle=0]{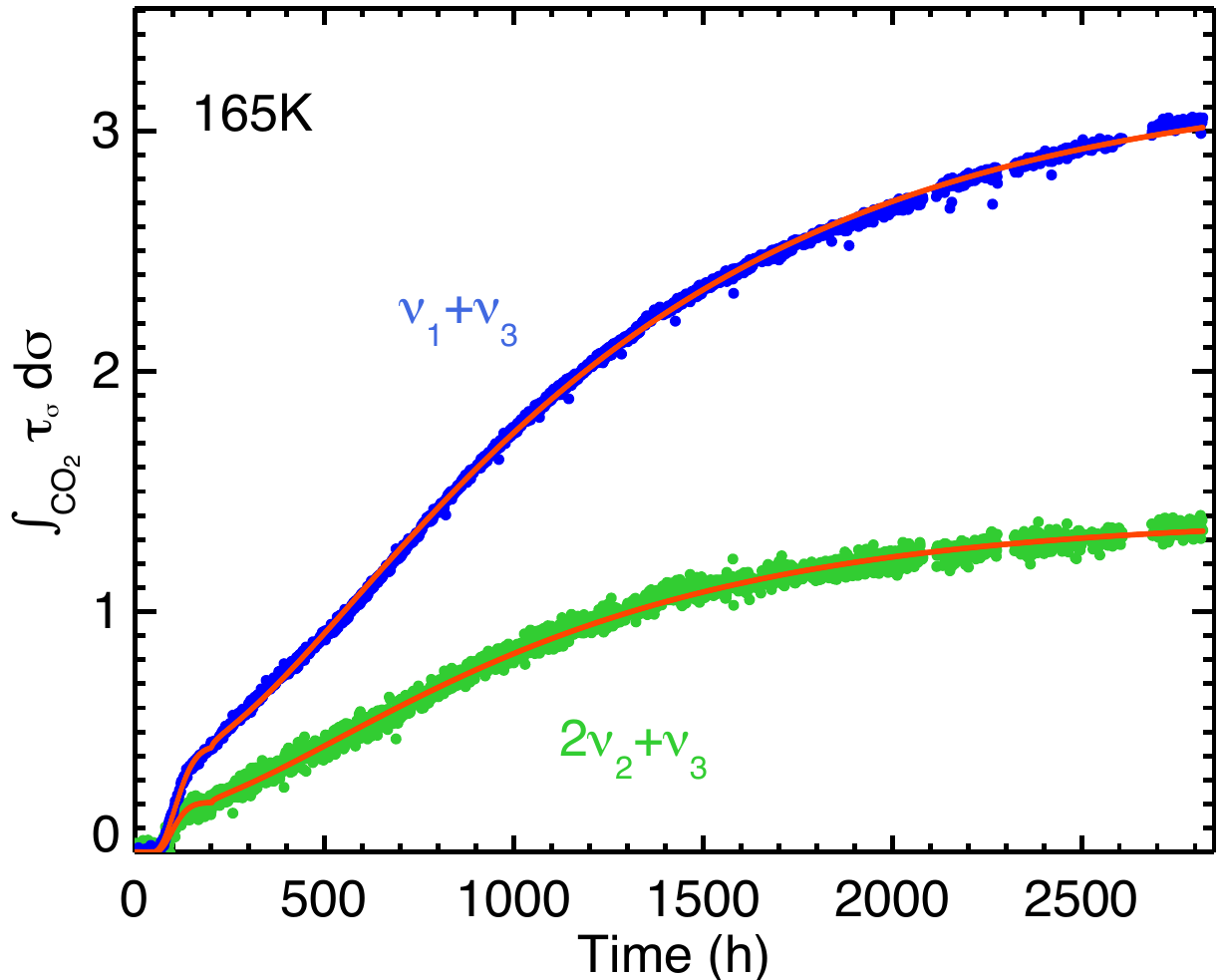}
\includegraphics[width=0.49\columnwidth,angle=0]{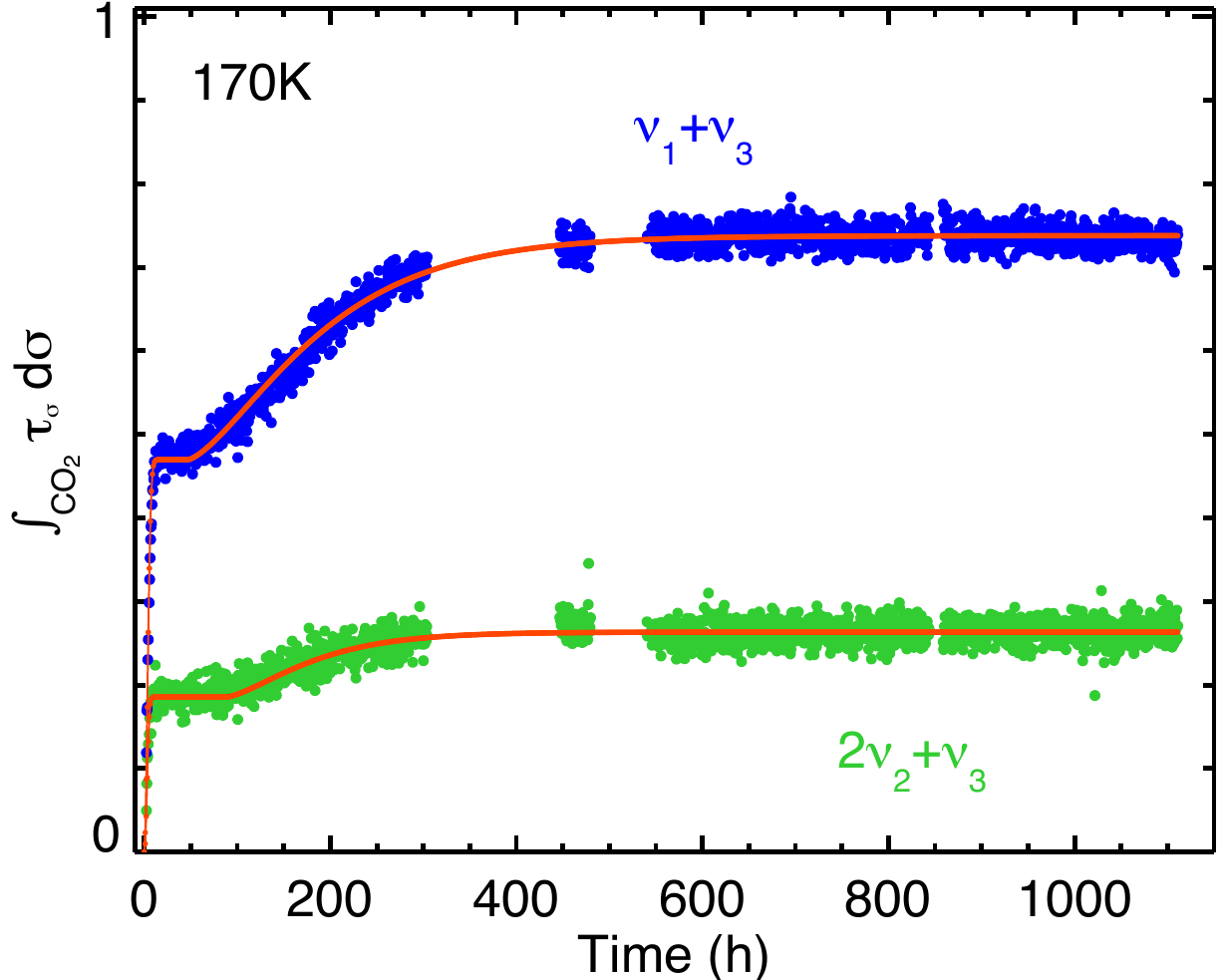}
\includegraphics[width=0.49\columnwidth,angle=0]{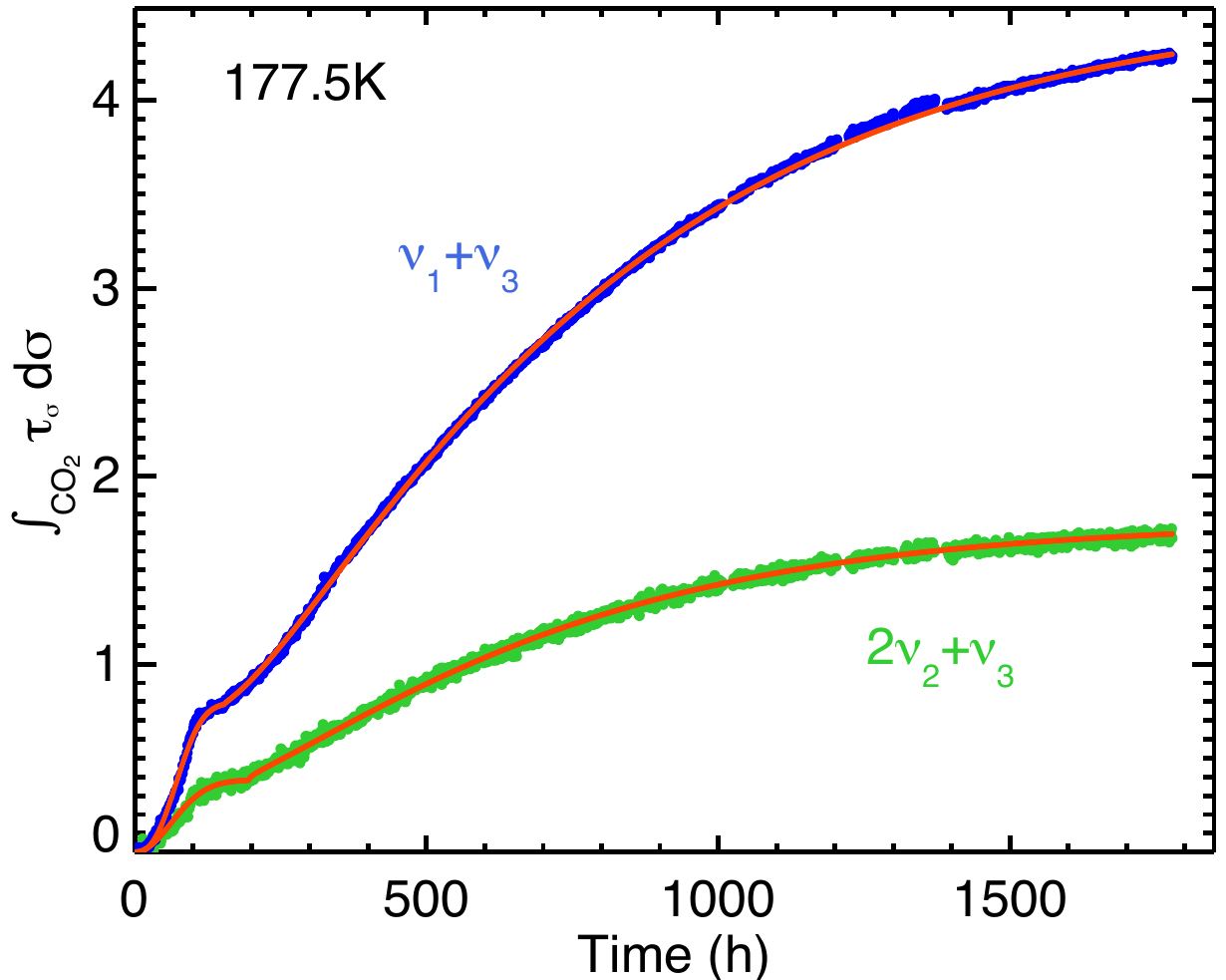}
\includegraphics[width=0.49\columnwidth,angle=0]{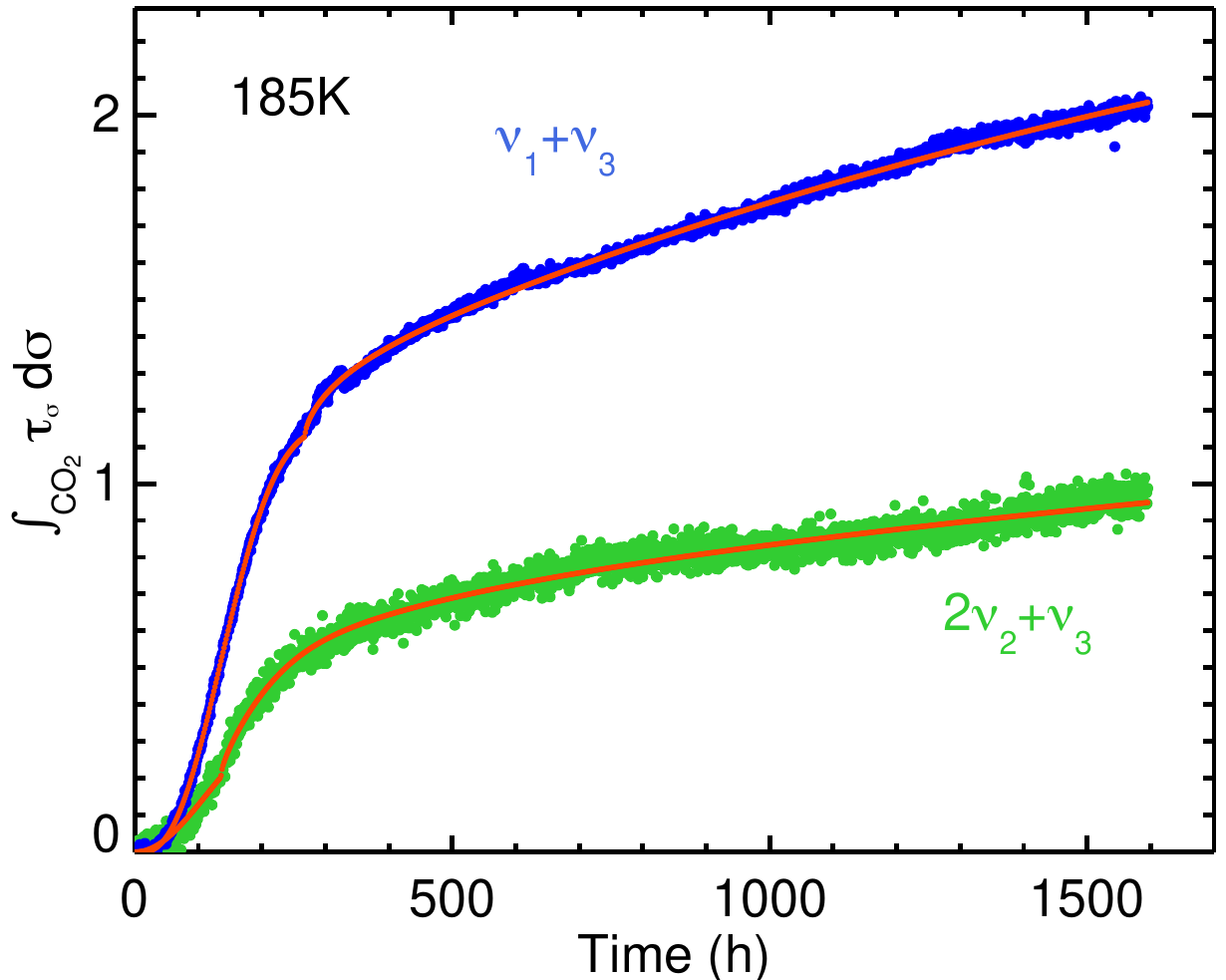}
\includegraphics[width=0.49\columnwidth,angle=0]{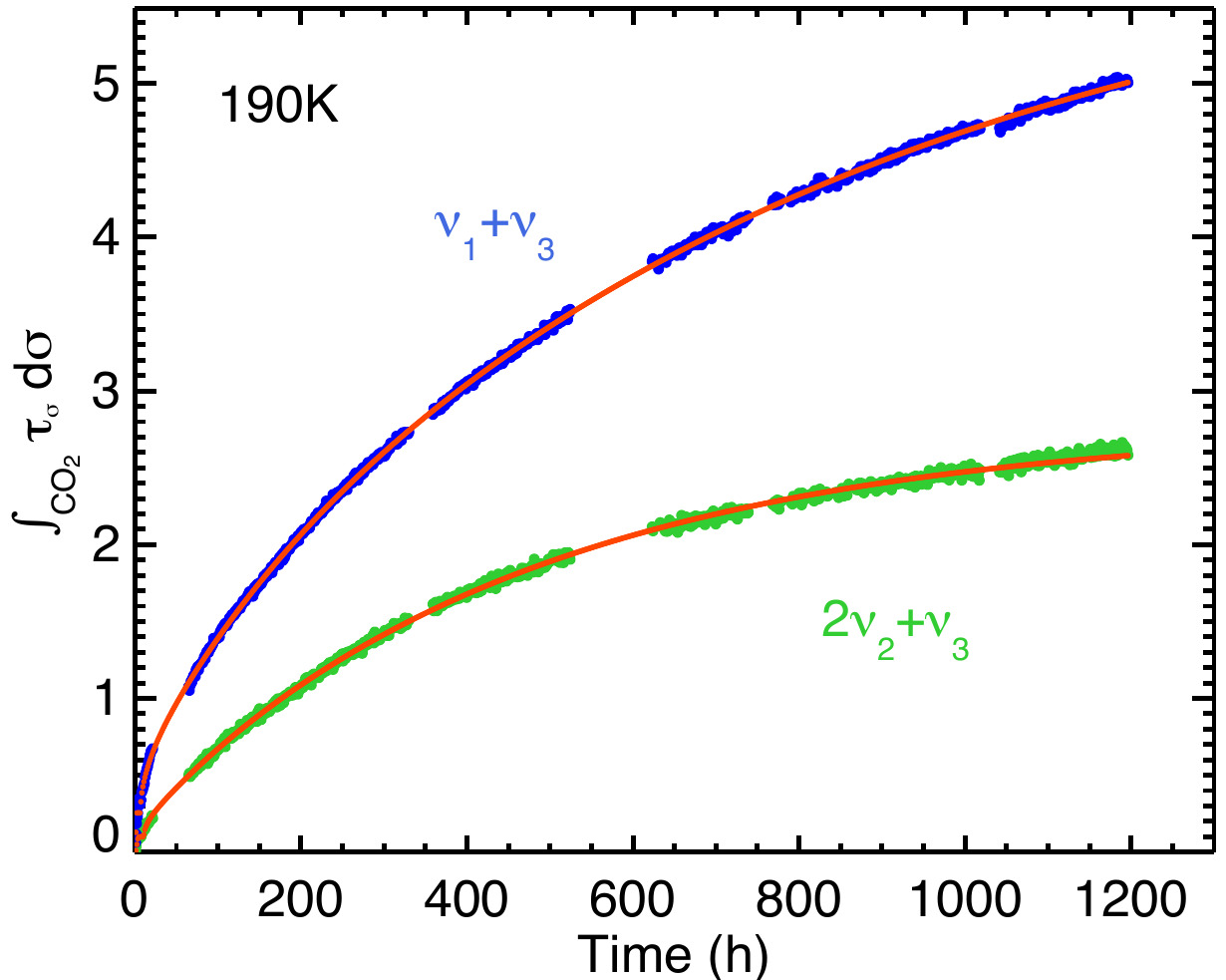}
\includegraphics[width=0.49\columnwidth,angle=0]{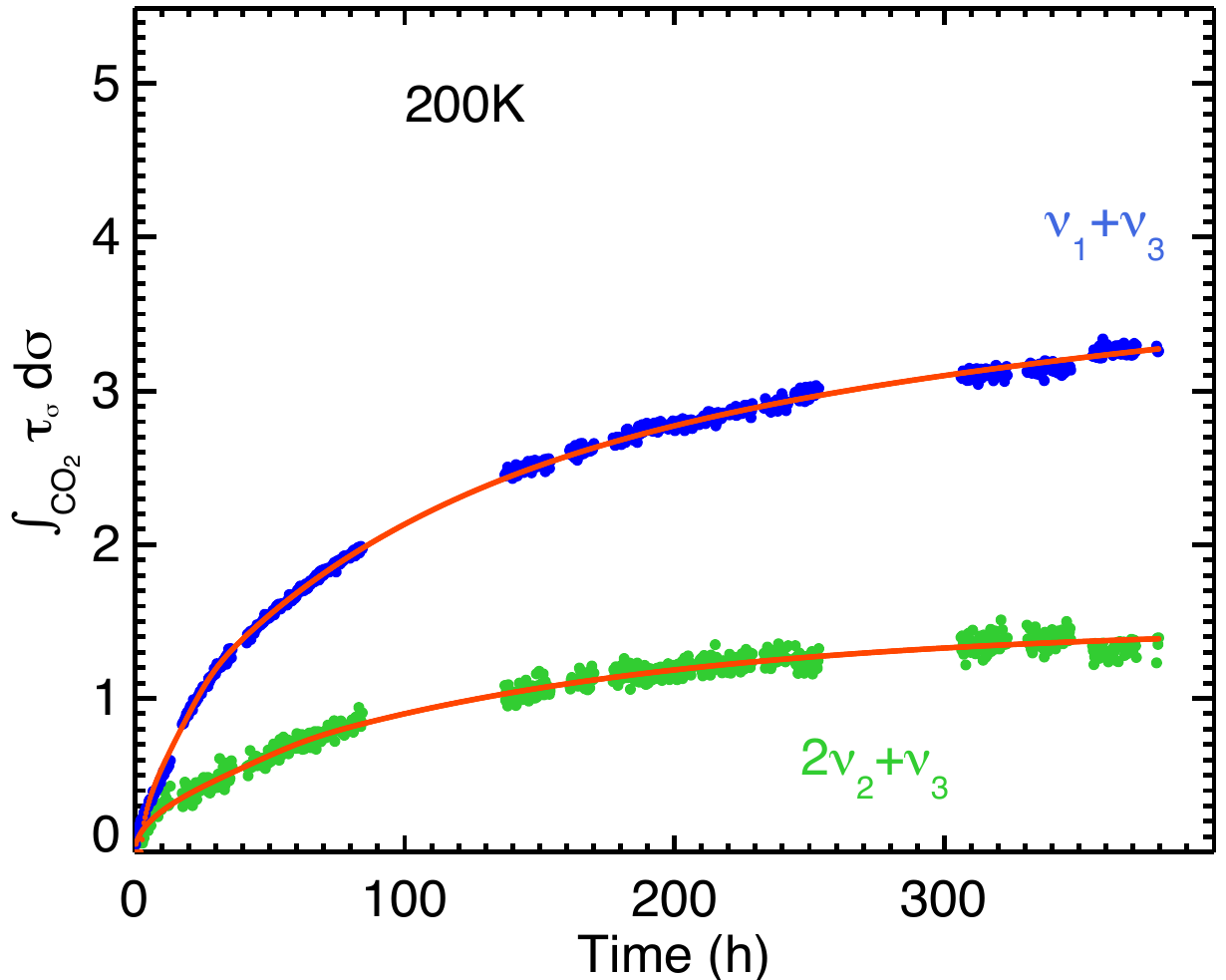}
\includegraphics[width=0.49\columnwidth,angle=0]{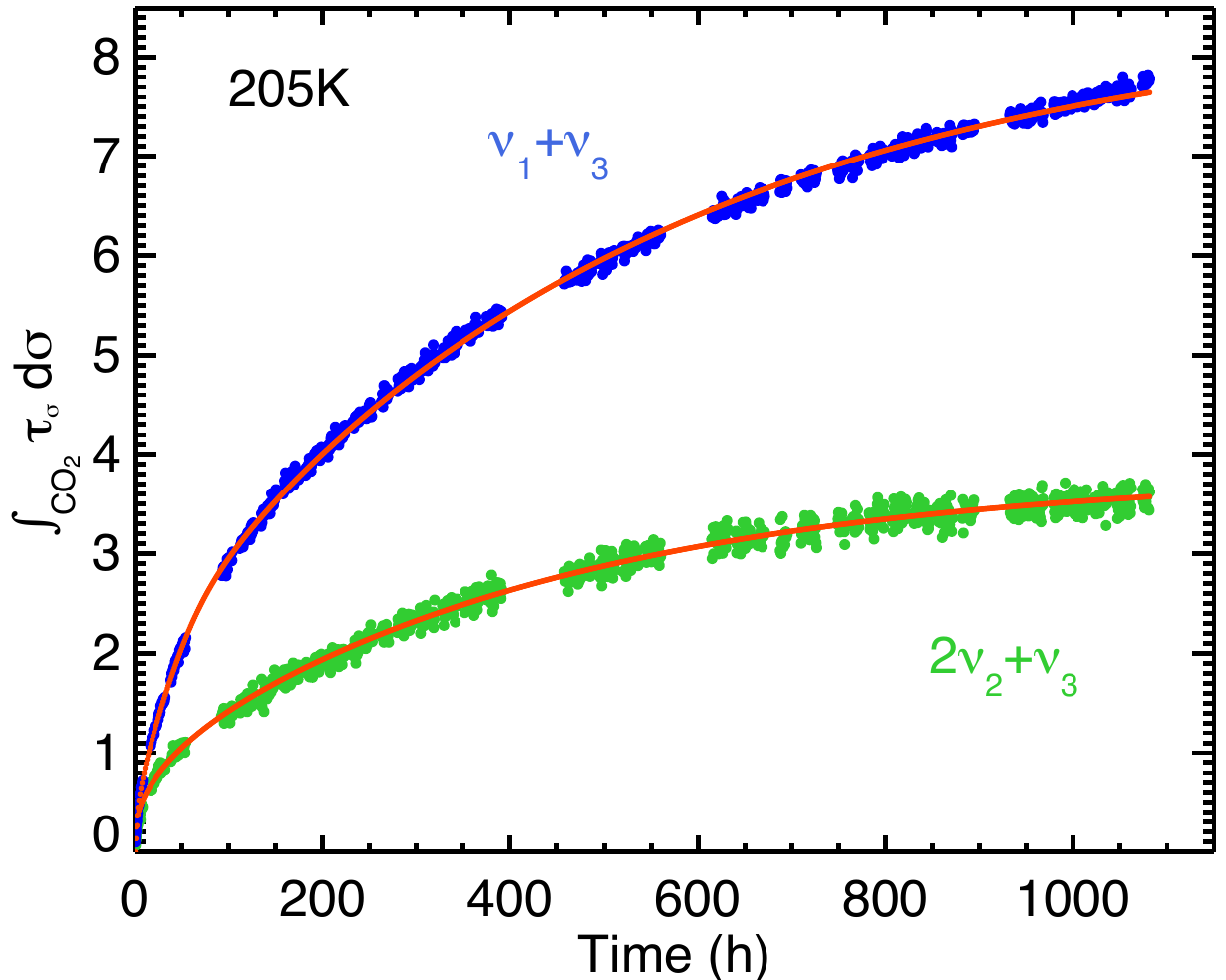}
\includegraphics[width=0.49\columnwidth,angle=0]{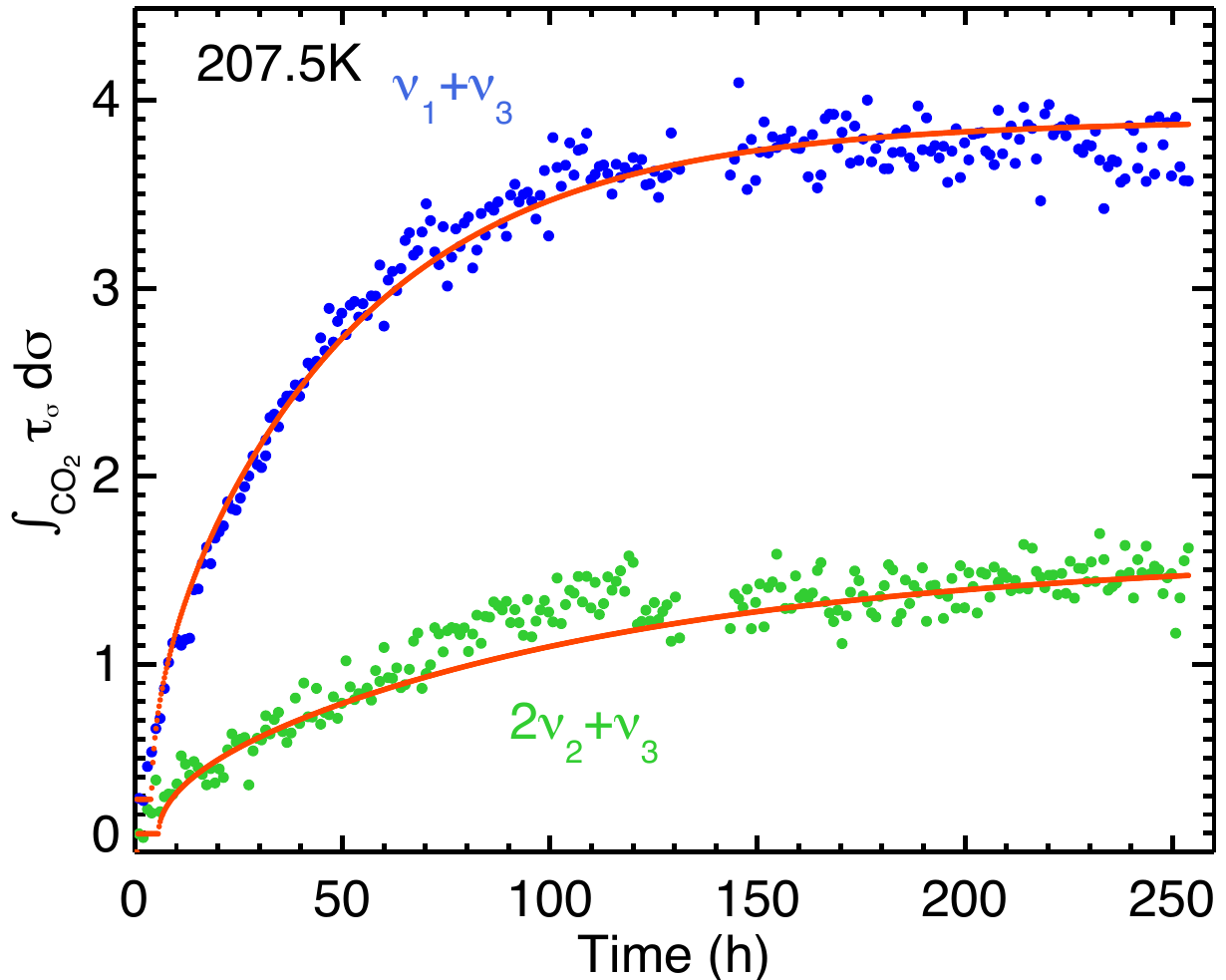}
\includegraphics[width=0.49\columnwidth,angle=0]{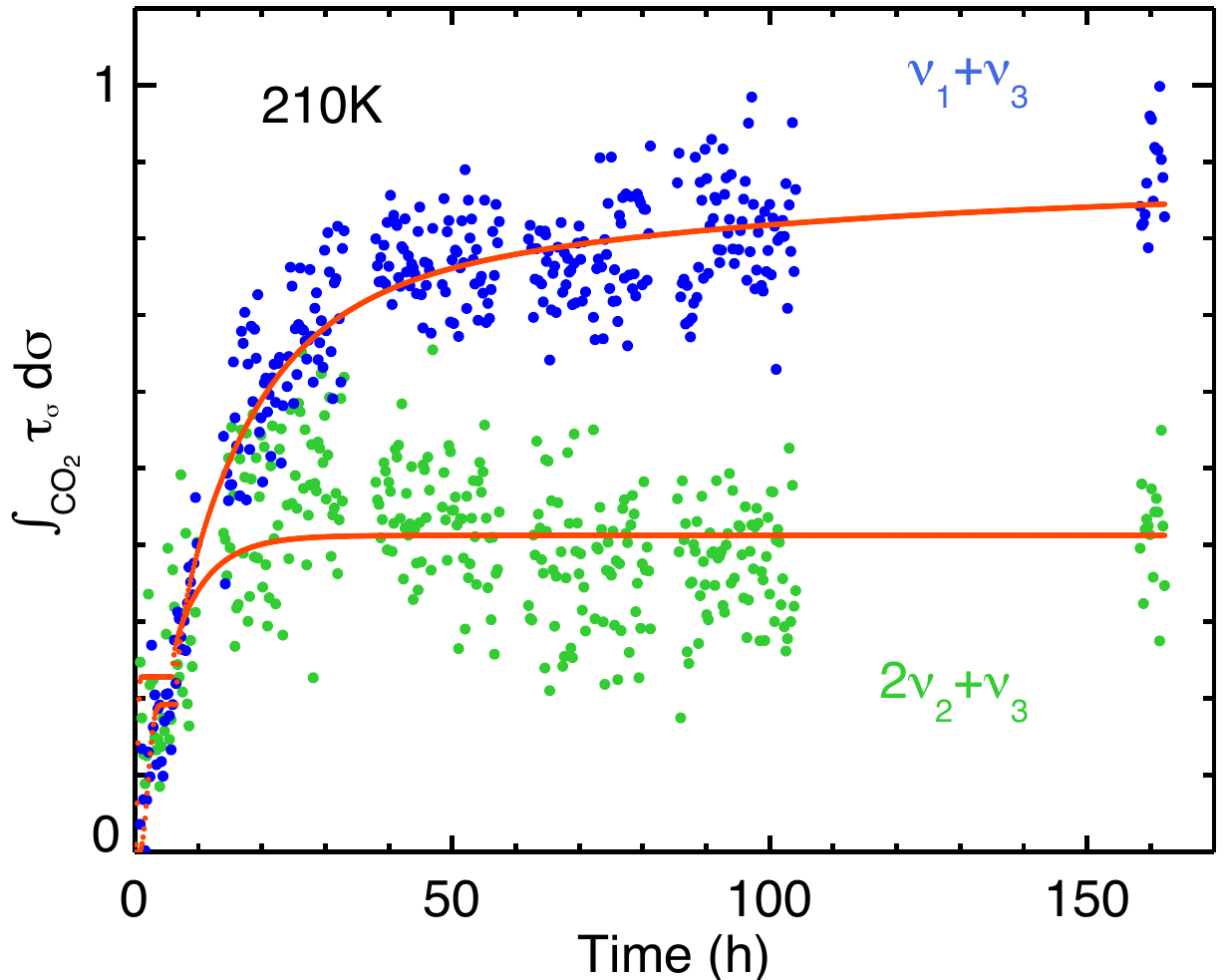}
\caption{Time-dependent evolution of the integrated optical depth of CO$_2$ clathrate infrared signatures in the various experiments performed. The $\rm \nu_1+\nu_3$ mode ($\sim$3700~$\rm cm^{-1}$) is represented in blue. The $\rm 2\nu_2+\nu_3$ mode ($\sim$3590~$\rm cm^{-1}$) is represented in green.}
\label{clathrations}
\end{center}
\end{figure}
%
%
\section{Results and astrophysical implications}
\label{results}
The time-dependent evolution of the integrated absorption of the $\rm \nu_1+\nu_3$ and $\rm 2\nu_2+\nu_3$ modes at temperatures from 165K to 210K, and with starting ice film thicknesses of several microns, is shown in Fig.~\ref{clathrations}. In the 155K experiment, the gas pressure is low enough to use the stretching $\rm \nu_3$ and bending $\rm \nu_2$ modes, but the constraint is only an upper limit because our experiment is limited to about 18 days.
The experiments are fitted with the model given by equation~\ref{equation_somme}. 
To find the best fits, the error function $\rm \chi^2=\sum_{i=1}^{N} (O_i-M_i)^2$, where $\rm O_i$ are the observed experimental point and $\rm M_i$ are the calculated model points, is minimised.
The results of the minimisation are over-plotted on the data shown in Fig.~\ref{clathrations} for the experiments performed, and the corresponding retrieved parameters are reported in Table~\ref{summary}.
In many practical cases, the diffusion-limited error bars on fitted diffusion models are overlooked, because the observed evolution of the process is smooth, and the numerous parameters in the models describing diffusion problems make it possible to have several neighbouring and almost equally good fits to the experimental data.
In order to retrieve the best-fit parameters and also estimate uncertainties (i.e. more systematic error bars) associated with the model, we use an amoeba algorithm, implementing a downhill simplex method \citep{Nelder2016}. The algorithm performs a multidimensional minimisation of the above error function. The initial parameters of the amoeba function are varied to explore the parameter space. The first moment of the solutions obtained for each parameter with different starting points are calculated. Error bars are given by the standard deviation for the solutions within three times the minimum value of the error function (i.e. a reduced $\rm \chi^2$). 
The JMAK equation~\ref{JMAK} parameters are retrieved with the best significance from the low-temperature experiments, where the kinetics do not overly mix the surface and volume diffusion-integrated effects.
In our experiments, the clathrate formation process is controlled by the diffusion rate of CO$_2$ molecules through the surface clathrate hydrate layer. 
The advantages of our technique over many previous experiments include the much simpler dimensionality of the system formed in our experiments, the possibility to vary the ice thickness, and the direct monitoring of the caged molecule signatures.
At low temperature, the surface nucleation (nucleation proceeding laterally from clathrate hydrate surface seeds) is expected to be more efficient than the mass transfer (permeation) through the surface followed by clathrate formation. 
The ice surface clathrate nucleation is limited to a depth of the order of a micron or to submicron depths in our explored range of temperatures, and therefore will  only penetrate and nucleate deeper by diffusion. 

\begin{figure}[htbp]
\begin{center}
\includegraphics[width=\columnwidth,angle=0]{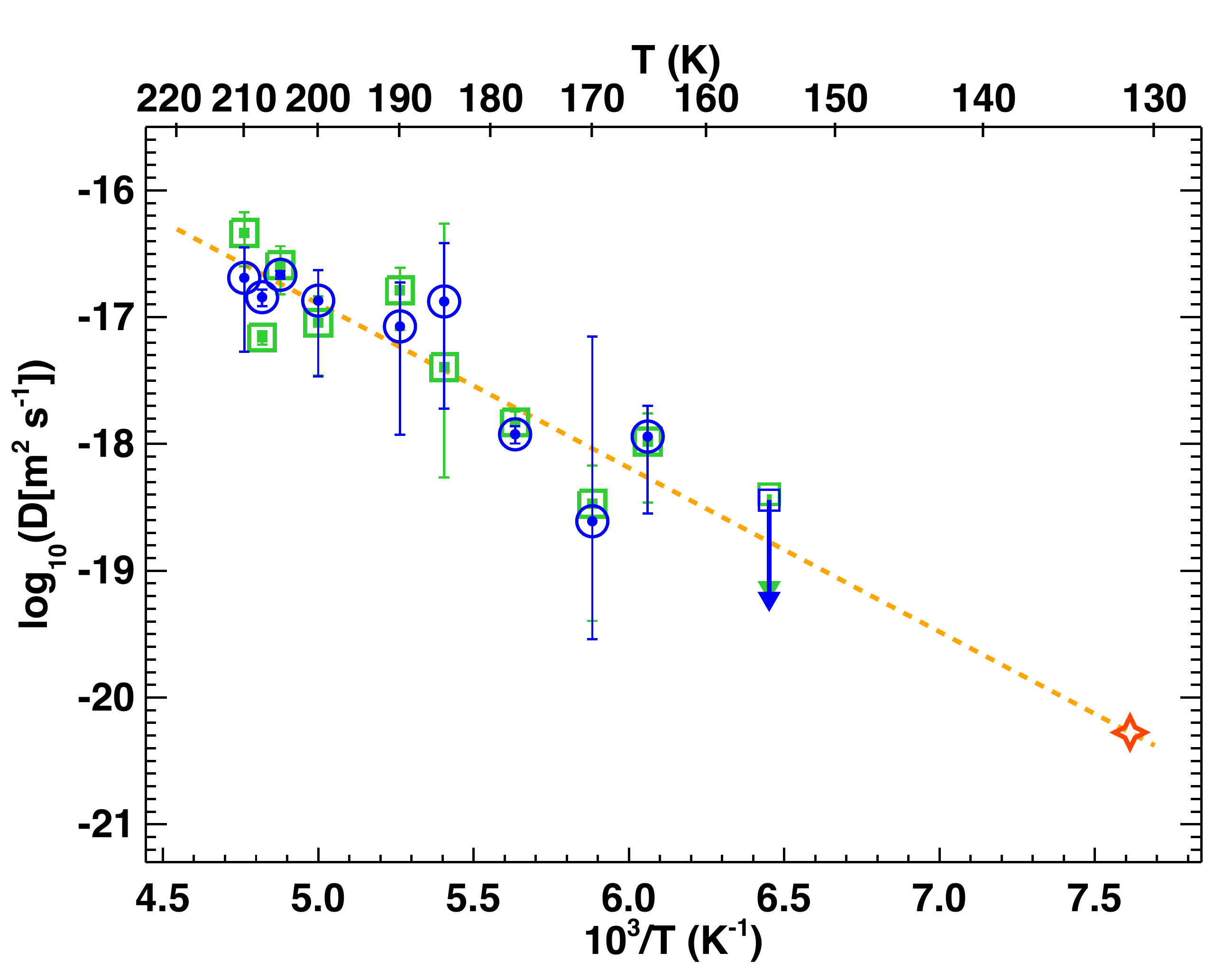}
\caption{Diffusion coefficient retrieved from the analyses of the time-dependent FTIR spectra of the $\rm 2\nu_2+\nu_3$ (blue circle) and $\rm \nu_1+\nu_3$ (green square) CO$_2$ caged in the clathrate hydrate (for the 155K point the $\nu_2$ and $\nu_3$ modes were used instead; see text). The data are fitted with an Arrhenius law shown by the dashed orange line. The red star indicates the expected temperature of the crossing point between the sublimation curve of pure CO$_2$ and the stability curve of the CO$_2$ clathrate hydrate below which the pure CO$_2$ solid phase dominates the stability.}
\label{coefficient_diffusion}
\end{center}
\end{figure}

The diffusion coefficient obtained from the fits are reported on a logarithmic scale as a function of the inverse of the temperature in Fig.~\ref{coefficient_diffusion}, with the error bars discussed above. We performed a fit to derive the apparent activation energy $\rm E_a$ corresponding to the temperature dependence of the diffusion, which approximately follows an Arrhenius law ($\rm D(T)\propto exp(-E_a / kT)$). 
The temperature dependence of the diffusion-controlled clathration process we measure in the 210-155K range possesses an activation energy of $\rm E_a = 24.7\pm{9.7} kJ/mol$.
This diffusion activation energy is lower than the one measured at temperatures above about 220K by \cite{Falenty2013} of $\rm E_a \approx 46 kJ/mol$, whereas \cite{Henning2000} measured an activation energy of $\rm E_a = 27.2 kJ/mol$ measured in the 230-263K range, both derived from neutron experiments on clathrates with a D$_2$O lattice.
Our determined value is in agreement with the <225K \cite{Falenty2013} activation energy of $\rm E_a = 19 kJ/mol$ based on experiments with an H$_2$O lattice; the error bar for these latter results was not specified but should largely overlap with our determination. 
If we include the \cite{Falenty2013} data below 225K, we find $\rm E_a = 22.6\pm{5.4} kJ/mol$.
Other studies of other clathrate hydrates led to somewhat comparable values of activation energy: C$_2$H$_6$ clathrates formed from ice + liquid C$_2$H$_6$$\sim$15 kJ/mol \citep{Vu2020}; 
Tetrahydrofuran clathrates formed under vacuum conditions $\sim$23 kJ/mol \citep{Ghosh2019}.
Most of these experimental results, with similar values for the apparent activation energy, suggest the view of a diffusion-limited process driven by the movement or reorganisation of water ice.
In their molecular dynamics models, \citet{Liang2016} found a high activation energy of $\rm E_a = 44\pm{6} kJ/mol$. 
In a water-solvent-driven diffusion of CO$_2$ in 
low density amorphous (LDA) ice, \cite{Ghesquiere2015} calculate an activation energy of 25$\pm$2.4 kJ/mol, in relatively good agreement with what we deduce at low temperature.

The nucleation of the seeds and the self-diffusion of water on the surface of ice grains could play an important role in the rearrangement of ice Ih into the first clathrate cage structure, with an activation energy on the order of 23 kJ/mol \citep{Nasello2007, Oxley2006}. Bulk (self-)diffusion of water into Ih single crystal ice multilayers of various thicknesses \citep{Brown1996}, and with other molecules \citep{Livingston2002}, proceeds with a much higher  activation
energy ($\rm E_a \approx 50-70 kJ/mol$). As stated in these latter articles, and especially for bulk ice, the ice morphology (amorphous, polycrystalline, or crystalline) will influence the efficiency of the diffusion into the ice, and the apparent activation energy.

%
%
%
Carbon dioxide is present in the ice mantles in interstellar molecular clouds with an abundance of several tens of percent \citep[e.g.][]{Boogert2015, deGraauw1996, Pontoppidan2008, Dartois2005}. This is probably not the place to look for clathrates formed by bulk diffusion of CO$_2$ into ice first. Indeed, consider a simplistic back-of-the-envelope calculation based on our findings, taking the lifetime of a dense cloud containing ices of the order of a few tens of millions of years (e.g. $\rm t_0\approx5\times10^7$ years) as the maximum diffusion time. To nucleate a $\rm r=0.1\mu$m ice grain clathrate by bulk diffusion during this lifetime requires $\rm D\approx r^2/t_0=6\times10^{-30} m^2.s^{-1}$. 
Extrapolating the diffusion coefficient from Fig.~\ref{coefficient_diffusion} to lower temperatures would require maintaining the icy grain exposed to CO$_2$ at temperatures higher than about 65K, which is well above typical ice mantle temperatures.
A clathrate could alternatively form by restructuring a previously condensed simple hydrate CO$_2$:H$_2$O mixture during a hotter phase. However, the driving force forming a crystalline compound phase such as a clathrate hydrate at such low temperatures before water ice sublimation raises many questions. The transformation kinetics associated with such low temperatures experienced by ice mantle grains still needs to be better understood.

Clathrate hydrates may be more relevant to protoplanetary disks.
In models of the early protosolar solar nebula (PSN), the protosolar disk pressure and temperature (P,T) follow a so-called cooling curve.
This curve can be associated to a thermodynamic path for the existence of clathrate hydrates \citep[e.g.][]{Mousis2010}. However, if satisfying the thermodynamics equilibrium curves is a necessary condition for the existence of a clathrate hydrate phase, it is a far too simplistic assumption to prescribe that clathrate hydrates are formed as soon as the P,T stability conditions are met. The composition range, kinetics, and competing phase formation are also essential parameters determining the existence of clathrate hydrates. If during the PSN evolution, water ice condenses first onto small- to intermediate-sized bodies (i.e. not the ones forming planets and/or large satellites where pressure will make the difference), because of a progressive decrease of the temperature of the PSN, this will limit the scope of direct clathrate nucleation in such objects. Unless exposed to moderate pressures for a long time, kinetics and thermodynamics favour CO$_2$ and H$_2$O separate sequential ice phase condensation, or eventually low-temperature hydrate (non-clathrate) mixing.

CO$_2$ has been directly observed in the gas phase in many comets or inferred from the CO Cameron excited band resulting from the photodestruction of CO$_2$ molecules \citep[e.g.][]{Fougere2016, Bockelee2016, Steckloff2016, Migliorini2016, McKay2016, McKay2015, Ootsubo2012, AHearn2012, Feaga2007, Colangeli1999, Crovisier1997, Feldman1997, Weaver1994, Combes1988}, raising the possibility of
 the presence of CO$_2$ clathrate hydrate.
Clathrate hydrates can, in particular, be a posterior to formation phenomenon occurring in comets, particularly when mobility and internal pressures build up during the comet entry into the inner Solar System, possibly forming a subsurface clathrate layer, including the CO$_2$ layer in comets \citep[e.g.][]{Marboeuf2011}, which will consequently act as a permeation barrier or disrupt the degassing. Cometary models include the clathrate phase, among others (crystalline, amorphous, pure condensed phases, trapped hydrates), to explore their influence \citep[e.g.][]{Marboeuf2014, Brugger2016}. 
Observationally, if clathrate hydrates are shown to be present in comets (i.e. detected by direct techniques such as the observation of spectroscopic signatures), then they will have to be taken into account when modelling the degassing pattern of species in comets through the physical properties of this specific phase, such as the kinetics, permeation filter, and/or their destabilisation.
Direct observational proof of clathrate hydrates remains elusive in these Solar System objects.

The case of planets, satellites, or any astrophysical body where significant pressures can build up are the objects that are most prone to hosting clathrate hydrates.
The Jovian Europa, Ganymede, and Callisto, Saturn's Iapetus, Phoebe, and Hyperion, the Uranian Ariel, Umbriel, and Titania, and Neptune's Triton \citep[e.g.][]{Brown2006, Buratti2005, Clark2005, Hibbitts2000, Hibbitts2003, Quirico1999} have all been shown to harbour CO$_2$. 
The stability of clathrates at depth in several objects has been mentioned in numerous studies \citep[e.g.][]{Kamata2019, Choukroun2013, Castillo-Rogez2012, Tobie2012, Sohl2010, Hand2006, Prieto-Ballesteros2005, Loveday2001, Stevenson1986, Miller1985}. 
These studies mention that from a thermodynamical point of view, clathrate hydrates are stable at depth, potentially mobilising from `thin' ice layers of a few kilometres in size up to hundreds of kms, and that clathrate hydrates are likely stable close to the surface, for example in the crust and the possible ocean of Europa, potentially in ice caps (seasonally) on Mars, or from the very surface for cold objects such as Titan.
Clathrate hydrates are likely to be present in the subsurface, affecting its mechanical and fluid behaviour, and would be stable, if present, at the surface of many very cold natural satellites, or objects orbiting in the outer regions of the Solar System. However, up to now, observed spectroscopic signatures do not show evidence for the formation of CO$_2$ clathrate at the surface.

Specifically, Mars has prompted some discussion about CO$_2$ clathrate hydrates. Part of the martian surface might experience temperature and pressures that pass through the CO$_2$ clathrate stability equilibrium \citep[e.g.][]{Longhi2006}.  CO$_2$ is abundant in the atmosphere of Mars and in the solid phase there, with some CO$_2$ ice deposits layered with H$_2$O in the south pole \citep{Buhler2019, Titus2003} and some pure water ice phase in the north pole \citep{Grima2009} or just below the surface even at moderate to high latitudes \citep{Piqueux2019}. 
For a stable clathrate field, clathrate must be able to (re-)form and/or resist with kinetics compatible with the latitude-dependent martian daily,  annual and up to secular pressure and temperature variations
\citep[e.g.][]{Smith2016, Byrne2009, Nye2000}. For a diffusion coefficient of $\rm 10^{-18}-10^{-20} m^2.s^{-1}$, which, according to our measurements, corresponds to the 170K-135K range, that is, temperatures typical of high latitudes, the clathration of 10~$\rm\mu m$ of ice takes approximately 30 to 320 years, respectively.
It is noteworthy that the CO$_2$ sublimation curve boundary crosses the CO$_2$ clathrate hydrate stability curve. This implies that below the corresponding crossing-point temperature shown in Fig.~\ref{coefficient_diffusion} with an orange symbol (in the 130.2-132.5~K range as stated in \cite{Dartois2009}, including uncertainties in the different clathrate equilibrium experiments), the CO$_2$ clathrate is stable, but the kinetics of formation by diffusion of CO$_2$ in water ice will most probably be considerably hindered by the formation of pure solid CO$_2$ crusts or layers. Therefore, diffusion into the ice at the interface below this point will no longer proceed via the contact with gaseous CO$_2$. It is expected that a change of behaviour in the diffusion problem will occur. 
The presence of CO$_2$ clathrates is also discussed in the context of the ancient Mars, and is though to have possibly influenced the past history of the martian climate evolution, at earlier times, where surface and subsurface conditions may have been more favourable to clathrate hydrate formation \citep[e.g.][]{Chassefiere2013}.
\section{Conclusion}

We measured the temperature-dependent diffusion-controlled formation of a carbon dioxide clathrate hydrate in the 155-210~K range by direct monitoring of the caged CO$_2$ specific clathrate hydrate infrared signatures.
The diffusion coefficient associated with the clathrate diffusion-controlled nucleation kinetics was deduced. It proceeds with an activation energy of $\rm E_a = 24.7\pm{9.7} kJ/mol$,
which is lower than the one found at higher ice temperatures.
Combining these measurements with previous kinetics measurements, the data are now available to implement the kinetics of CO$_2$ hydrate formation from ice in models, for temperatures ranging from the clathrate--solid CO$_2$ crossing point around 130K, by slight extrapolation, up to the water triple point.
Looking at the low-temperature-dependent diffusion coefficient associated with clathrate formation, carbon dioxide clathrate hydrate nucleation is principally expected in planets or satellites, and may build up with moderate pressures and higher temperatures from simple hydrates in comets as they enter the inner Solar System.
\begin{acknowledgements}
Part of the equipment used in this work has been financed by the French INSU-CNRS program 'Physique et Chimie du Milieu Interstellaire' (PCMI). We wish to acknowledge past years technical support on the experiment cell conception from A. Arondel, M. Bouzit, M. Chaigneau, N. Coron, B. Crane, T. Redon. 
\end{acknowledgements}

\end{document}